\documentclass[a4paper,11pt]{article}
\pdfoutput=1 

\usepackage{jheppub} 
\usepackage{booktabs}
\usepackage{todonotes}

\newcommand{\bbbar}{\ensuremath{\text{b}\bar{\text{b}}}}

\newcommand{\ttbar}{\ensuremath{\text{t}\bar{\text{t}}}}

\newcommand{\ttbarH}{\ensuremath{\text{t}\bar{\text{t}}}\text{H}}
\newcommand{\ttbarHbb}{\ensuremath{\text{t}\bar{\text{t}}\text{H}\text{(\bbbar)}}}

\newcommand{\Mttbar}{\ensuremath{M_{\ttbar}}}
\newcommand{\MttbarH}{\ensuremath{M_{\ttbar\text{H}}}}

\newcommand{\pT}{\ensuremath{p_\text{T}}}

\newcommand{\Sherpa}{Sherpa}

\newcommand{\muR}{\ensuremath{\mu_R}}
\newcommand{\muF}{\ensuremath{\mu_F}}

\newlength\cmsTabSkip

\newcommand{\Pythia}{\textsc{Pythia8}} 
\newcommand{\Powheg}{\textsc{Powheg}} 
\newcommand{\MGaMCatNLOOnly}{\textsc{MG5\_aMC@NLO}} 

\newcommand{\ab}{\ensuremath{\text{b}}}

\newcommand{\aZ}{\ensuremath{\text{Z}}}

\newcommand{\Vptmiss}{\ensuremath{{\vec p}_{\mathrm{T}}\hspace{-0.78em}/\kern0.45em}}

\newcommand{\invfb}{\ensuremath{\text{fb}^{-1}}}

\newcommand{\TeV}{\text{TeV}}

\newcommand{\GeV}{\ensuremath{\text{GeV}}}

\newcommand{\msbar}{\ensuremath{\overline{\text{MS}}}}
\newcommand{\alphas}{\ensuremath{\alpha_\text{s}}}

\newcommand{\ttZ}{\ensuremath{\text{t}\bar{\text{t}}\text{Z}}}

\newcommand{\ttgamma}{\ensuremath{\text{t}\bar{\text{t}}\gamma}*}

\newcommand{\ttbjets}{\ensuremath{\text{t}\bar{\text{t}}+\ensuremath{\ab}\text{-jets}}}

\newcommand{\PH}{\ensuremath{\text{H}}}

\newcommand{\Hbb}{\ensuremath{\PH\rightarrow\bbbar}}

\newcommand{\thirteenTeV}{\ensuremath{\ensuremath{\sqrt{s}=13}~\text{TeV}}}

\newcommand{\PoleMass}{\ensuremath{ m_{\text{t}}^{\text{pole}}}}
\newcommand{\MSbarMass}{\ensuremath{m({\muR})}}
\newcommand{\HiggsMass}{\ensuremath{m_\text{H}}}
\newcommand{\mTop}{\ensuremath{ m_{\text{t}}}}
\newcommand{\LambdaQCD}{\ensuremath{ \Lambda_{\text{QCD}}}}

\title{Cross-Sections for \ttbarH\ production with the \\ Top Quark \msbar\ Mass}

\author[a]{~Maria Aldaya Martin,}
\author[b]{~Sven-Olaf Moch,}
\author[a]{~Andrej Saibel}

\affiliation[a]{Deutsches Elektronen Synchrotron, \\Notkestraße 85, D-22607 Hamburg, Germany}
\affiliation[b]{II. Institut für Theoretische Physik, Universit\"at Hamburg\\
	Luruper Chaussee 149, D-22761 Hamburg, Germany}

\emailAdd{maria.aldaya@desy.de}
\emailAdd{sven-olaf.moch@desy.de}
\emailAdd{andrej.saibel@desy.de}

\abstract{
We study the impact of the top quark mass renormalized in the \msbar\ scheme on the \ttbarH\ production cross-sections as an alternative to theory predictions with the conventionally used pole mass scheme.
The differential cross-sections at next-to-leading order in perturbative QCD with stable top quarks show a moderate decrease in scale uncertainties for the \msbar\ mass renormalization scheme compared to the on-shell one.
The shape of the differential distributions is not affected much, the largest differences being observed in the invariant mass distributions of the \ttbar\ and the \ttbarH\ systems.}

\begin{document} 
\maketitle
\flushbottom

\section{Introduction}
\label{sec:introduction}

A Higgs-like particle was observed by the ATLAS~\cite{Aad:2012tfa} and 
CMS~\cite{Chatrchyan:2012ufa} collaborations at the Large Hadron Collider (LHC) in the year 2012.
To confirm that this particle with spin 0 and mass $\HiggsMass=125.10\pm0.14$ \GeV\ \cite{PDG2020} is indeed the Higgs boson of the Standard Model (SM), 
studies of all properties of the newly discovered particle need to be conducted.
To achieve this goal, all production and decay modes of the Higgs-like particle are investigated. 
Especially the study of the Yukawa coupling is of great interest, since the observation of the Higgs-Yukawa coupling confirms the prediction of the SM, in which the masses of the fermions are generated by interactions with the Higgs field.
Conversely, significant deviations from the SM predicted cross-sections would be an indicator for new physics beyond the SM.

Since the top quark is by far the heaviest known elementary particle and the strength of the Yukawa coupling is proportional to the mass of the fermions, the searches for physics processes involving the production of top quark pairs with an associated Higgs boson (\ttbarH) are most promising for the discovery of the coupling. 
The observation of \ttbarH\ production and, therefore, the confirmation of the existence of the top-Higgs Yukawa coupling was claimed by both ATLAS~\cite{bib:ObservationttHATLAS} and 
CMS~\cite{bib:ObservationttHCMS} in the year 2018.
This achievement was made possible by a statistical combination of the results of the data analyses for the \ttbarH\ production with different channels of the Higgs boson decay, i.e., to di-bosons, $\tau^+\tau^-$, $\gamma\gamma$, and \bbbar\ -pairs.

The uncertainties originating from systematic effects in the combined measurements of both collaborations surpass the statistical uncertainties arising from the limited size of the recorded data samples already now. 
Looking further into the future, the discovery of \ttbarH\ leads to an era of precise measurements probing the coupling with larger data sets and improved analysis techniques.
In particular, the High Luminosity LHC (HL-LHC)~\cite{ApollinariG.:2017ojx} will allow for precise differential cross-section measurements of \ttbarH\ production.
Systematic uncertainties of the \ttbarH\ measurements originating from the uncertainties on the theory predictions will become increasingly important in the future. 
This motivates studies which aim at quantifying and reducing these uncertainties.

Currently available theory calculations for the \ttbarH\ production process at the LHC are based on the QCD radiative corrections to next-to-leading order (NLO), which are known since long~\cite{Beenakker:2001rj,Reina:2001sf,Dawson:2003zu} and have been matched to Monte Carlo parton shower generators such as in the \Powheg+\Pythia\ framework for \ttbarH\ production~\cite{Hartanto:2015uka} that is used by ATLAS and CMS for the simulation of the signal process.\footnote{
The most commonly used implementations of NLO QCD corrections matched to Monte Carlo parton shower generators are 
\Powheg~\cite{Hartanto:2015uka}, HELAC-NLO~\cite{Garzelli:2011vp}, 
\MGaMCatNLOOnly~\cite{Alwall:2014hca} and 
\Sherpa\ \cite{Sherpa:2019gpd,Buccioni:2019sur}, 
with \Powheg\ being preferred due to its small amount of events with negative weights.}  
These QCD theory predictions have been improved by performing resummations of large threshold logarithms~\cite{Kulesza:2015vda,Kulesza:2017ukk,Broggio:2015lya,Broggio:2016lfj} 
to next-to-next-to-leading (NNLL) accuracy and of Coulomb corrections ~\cite{Ju:2019lwp}. 
In addition, also the effect of off-shell top quarks and the NLO electroweak corrections have been studied~\cite{Frixione:2014qaa,Zhang:2014gcy,Frixione:2015zaa,Denner:2015yca,Denner:2016wet,Stremmer:2021bnk}.
Recently, also the next-to-next-to-order (NNLO) QCD corrections for the flavor off-diagonal partonic channels in \ttbarH\ production became available~\cite{Catani:2021cbl}.

While these efforts help to reduce the uncertainties from the truncation of the perturbative expansion, which are conventionally studied by scale variations, 
it is also essential to investigate uncertainties related to definition and the choice of input parameters in the theory predictions, such as the top quark mass.
Currently, the \ttbarH\ predictions are using exclusively the on-shell mass renormalization scheme for the top quark mass.\footnote{
The Monte Carlo simulations used in data analyses for signal modeling compute the \ttbarH\ production process at the NLO precision with the pole mass \PoleMass.} 
This motivates us to consider different mass renormalization schemes, in the particular the so-called running mass in the \msbar\ scheme~\cite{Saibel:2021the}.
Such a study is of interest by itself, but also important in order to confirm 
that no major systematic effect was forgotten in the experimental analyses, which have contributed to the observation of the \ttbarH\ process.

In this letter we investigate the impact of the running mass on the differential cross-sections in \ttbarH\ production at NLO in QCD. 
In Sec.~\ref{ch:PhenomenologyttH:sec:Calculation} we set up the theory framework and discuss some technical details of the computation. 
In Sec.~\ref{ch:PhenomenologyttH:sec:Results} we present results and a brief phenomenological study and we conclude in Sec.~\ref{ch:PhenomenologyttH:sec:Summary}.
Numerical results for the differential cross-sections are listed App. \ref{app:RunningCrossSection}.


\section{Theoretical framework}
\label{ch:PhenomenologyttH:sec:Calculation}

The on-shell mass of a heavy quark, defined through the pole of the propagator, is a well-defined concept in perturbative quantum field theory, but it has its disadvantages.
In the pole mass scheme, the particles are assumed to be free, asymptotic states, an assumption not well justified for quarks in nature due to confinement.
Furthermore, the pole mass exhibits poor convergence in the perturbative expansion, having an intrinsic uncertainty in the order of \LambdaQCD\ \cite{Beneke:1994sw,Bigi:1994em}.

Short-distance mass, such as heavy quark masses renormalized in the \msbar\ scheme do not have such limitations and can, thus, be used for predictions of differential cross-sections to reduce uncertainties.
For \ttbar~cross-sections, both total \cite{Langenfeld:2009wd} and differential \cite{Dowling:2013baa,Garzelli:2020fmd}, it has been shown that the \msbar\ renormalization scheme for the top quark mass improves  the apparent convergence of the perturbative expansion along with a reduced scale dependence. These improvements lead to smaller total theory uncertainties and, therefore, to predictions which are desirable for the usage in cross-section measurements and comparisons of the experimental results to theory.

The computation of differential cross-sections for \ttbarH\ production
proceeds in analogy to the \ttbar\ case, see e.g.,~\cite{Langenfeld:2009wd}.
Starting from an expansion in powers of the strong coupling $\alphas$ 
up to NLO in QCD, 
the (differential) cross-section with the on-shell mass $\PoleMass$
can be written as
\begin{align}
\label{eq:DiffXSPoleMass}
\frac{d \sigma\left(\PoleMass\right)}{d X}
=\left(\frac{\alphas}{\pi}\right)^{2}\, 
\frac{d \sigma^{(0)}\left(\PoleMass\right)}{d X}
+\left(\frac{\alphas}{\pi}\right)^{3}\, 
\frac{d \sigma^{(1)}\left(\PoleMass\right)}{d X}
+O\left(\alphas^{2}\right)
\, ,
\end{align}
where $\sigma^{(0)}$ is the Born cross-section, $\sigma^{(1)}$ the NLO contribution, and $X$ the observable in which the differential cross-section is calculated.

The conversion of the \ttbarH\ production cross-sections in the pole mass renormalization scheme to the \msbar\ mass one proceeds as follows. 
The on-shell top quark mass \PoleMass is rewritten in terms of the \msbar\ mass \MSbarMass\ using standard formulae
\begin{align}
\label{eq:MassRelation}
\PoleMass = \MSbarMass \left(1 + \frac{\alphas}{\pi}d_1
+O\left(\alphas^{2}\right) \right)
\, ,
\end{align}
where the perturbative expansion in \alphas\ on the right hand side is truncated at NLO for brevity.
Currently, the conversion in Eq.~(\ref{eq:MassRelation}) is known to 
four-loop order~\cite{Gray:1990yh,Melnikov:2000qh,Chetyrkin:1999qi,Marquard:2016dcn}. The NLO coefficient reads
\begin{align}
d_1 = \frac{4}{3}+\ln\left(\frac{\muR^2}{\MSbarMass^2}\right)
\, .
\end{align}
The differential cross-section in the \msbar\ scheme is then 
obtained by inserting the relation between the pole and the \msbar\ masses 
in Eq.~(\ref{eq:MassRelation}) and expanding in \alphas\ (see e.g., \cite{Dowling:2013baa}),
\begin{align}
\label{eq:DiffXSmsbar}
\lefteqn{
\frac{d \sigma\left(\MSbarMass\right)}{d X}=
\left(\frac{\alpha_{s}}{\pi}\right)^{2} \frac{d \sigma^{(0)}\left(\MSbarMass\right)}{d X}
}\nonumber \\
&+\left(\frac{\alpha_{s}}{\pi}\right)^{3}\left\{\frac{d \sigma^{(1)}\left(\MSbarMass\right)}{d X}
+d_{1} \MSbarMass \frac{d}{d \mTop}\left.\left(\frac{d \sigma^{(0)}\left(\mTop\right)}{d X}\right)\right|_{\mTop=\MSbarMass}\right\}	
+O\left(\alpha_{s}^{2}\right) 
\, ,
\end{align}
where $\sigma^{(0)}$ and $\sigma^{(1)}$, i.e., the first and the second term, 
correspond to the ones in Eq.~(\ref{eq:DiffXSPoleMass}), only with \MSbarMass\ replacing \PoleMass. 
The mass derivative of the Born cross-section evaluated at  $\mTop=\MSbarMass$, i.e., the third term, 
has to be calculated separately in order to obtain the \ttbarH\ cross-section at NLO with the top quark mass in \msbar\ scheme. 
For a given differential cross-section this can be done numerically. 
To that end, we use \MGaMCatNLOOnly\ \cite{Alwall:2014hca} (version 2.6.5) 
to obtain the Born level differential cross-sections varying 
the top quark mass (and the Yukawa coupling accordingly) 
in a range between $\mTop=150$ \GeV\ and $\mTop=175$ \GeV\ in steps of $\Delta \mTop=0.5$ \GeV.

The calculations in our study~\cite{Saibel:2021the} are performed using the $\textsc{MMHT2014}$ PDF sets \cite{Harland-Lang:2014zoa} at NLO with $\alphas(m_\aZ)=0.118$.
The \msbar\ mass of the top quark at the scale $\mTop$ is chosen to be $\mTop(\mTop) = 163.2$ \GeV, while the corresponding pole mass is $\PoleMass=172.5$ \GeV.
The mass of the Higgs boson is chosen to be $\HiggsMass = 125.0$ \GeV. 
The running of \alphas\ and of the top quark mass \MSbarMass\ are calculated with five light flavors with the help of the 
RunDec program~\cite{Chetyrkin:2000yt,Schmidt:2012az}.
The nominal value for the \muR\ and the \muF\ scales in the calculations is chosen to be $\mu_0=2m_{\text{t}}+m_{\text{H}}$.
When the mass of the top quarks is varied for the calculation of the mass derivative, the scales are also adjusted.
The scale uncertainties are estimated through variations of $\muR, \muF \in [0.5\mu_0,\,2\mu_0]$ and the value of the top quark \msbar\ mass $\mTop(\mTop)$ is kept unchanged between the scale variations.

\begin{figure}[tp]
	\centering
	\includegraphics[width=.9\textwidth]{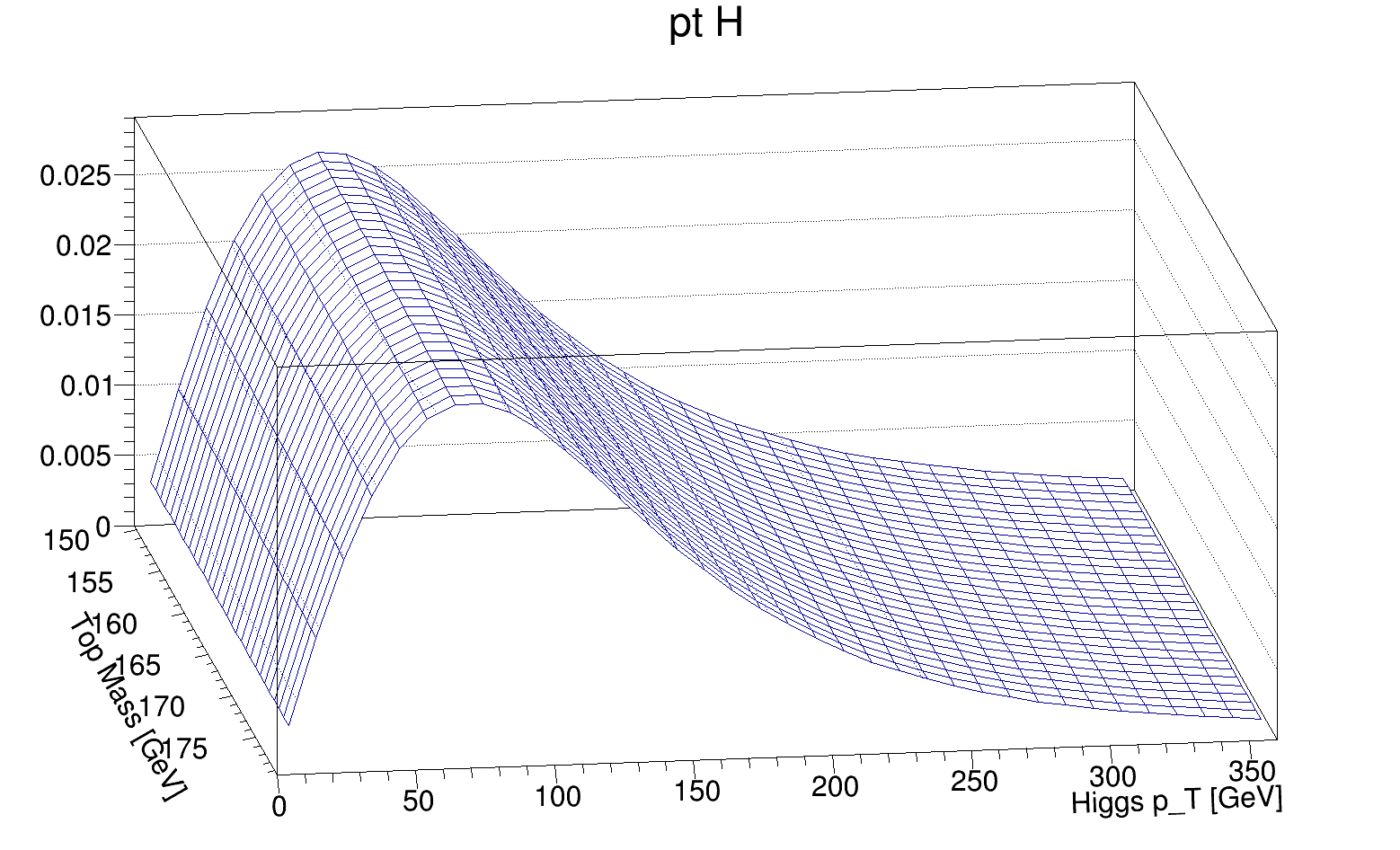}
	\caption{\label{fig:BornDerivative2d} 
Two dimensional histogram for the mass derivative of the Born-level \ttbarH\ cross-section as a function of the transverse momentum of the Higgs boson and the top quark mass in a range between $\mTop=150$ \GeV\ and $\mTop=175$ \GeV .}
\end{figure}
The calculation of the Born level differential cross-sections for \ttbarH\ production in Eq.~(\ref{eq:DiffXSmsbar}) in steps of $\Delta \mTop=0.5$ \GeV\ creates a two-dimensional distribution as a function of the observable $X$ and the mass of the top quark.
Fig.~\ref{fig:BornDerivative2d} illustrates this distribution for the 
\pT\ of the Higgs boson.
Since the mass derivative is estimated numerically, the Born cross-section as a function of \mTop\ 
is calculated with a statistical precision of $0.1\text{\textperthousand}$ 
on the inclusive cross-section from the Monte Carlo integration.
This ensures low enough statistical uncertainty of the mass derivative in each bin, which can then be obtained as 
\begin{align}
	\label{eq:derivative}
	\frac{d}{d \mTop}\left(\frac{d \sigma^{(0)}\left(\mTop\right)}{d X}\right) 
	\approx 
	\frac{\frac{d \sigma^{(0)}\left(\mTop + \Delta m\right)}{d X} 
	- \frac{d \sigma^{(0)}\left(\mTop\right)}{d X}}{\Delta m}
\, .
\end{align} 
For the NLO cross-sections in the pole mass scheme, i.e. 
the term $\sigma^{(1)}$ in Eq.~(\ref{eq:DiffXSmsbar}), 
a statistical uncertainty of $0.2\text{\textperthousand}$ on the total cross-section proves to be sufficient. 
With the mass derivative in Eq.~(\ref{eq:derivative}) all components 
are available to obtain the differential cross-section for \ttbarH\ production in the \msbar\ scheme in Eq.~(\ref{eq:DiffXSmsbar}). 
Compared to the required numerical precision of the cross-sections, the statistical uncertainties from the Monte Carlo integration 
on the mass derivative are negligible in the bulk of the distributions. 
They do become relevant, however, for the transverse momentum and invariant mass distributions 
in the region of $\gtrapprox 1$ \TeV.
Consequently, the binning of those distributions is chosen 
accordingly 
to minimize the statistical uncertainties compared to the scale uncertainties.
Despite being small, the statistical uncertainties of the mass derivative of the Born cross-section are added 
in quadrature to the scale uncertainties for consistency.


\section{Results and Discussion}
\label{ch:PhenomenologyttH:sec:Results}

Here we present the results for \ttbarH\ differential cross-sections 
with the top quark \msbar\ mass at NLO and compare them with calculations in the pole mass scheme with $\PoleMass=172.5$ \GeV.
We study the behavior of the top quark and the Higgs boson separately, as well as those of systems of particles.
Therefore, the distributions we study are 
the transverse momentum \pT\ and the rapidity $y$ of the top quarks 
and the Higgs boson, as well as the invariant mass of the top quark pair and the \ttbarH\ system.

The results are shown as the differential cross-sections, as well as 
two ratio graphs to facilitate comparisons.
The first ratio compares the shapes of the differential cross-sections 
in the two mass renormalization schemes.
For the second ratio, the scale uncertainties are normalized to the cross-section prediction in each bin for the two mass renormalization schemes.
Thus, the second ratio compares the relative scale uncertainties.
If the scale variations calculated in the interval 
$[0.5\mu_0,\, 2\mu_0]$ point in the same direction in a given bin, then the larger deviation from the nominal cross-section 
is taken as the uncertainty. 
The values in each bin of all differential cross-sections 
considered in this Section are listed in App.~\ref{app:RunningCrossSection}.

\begin{figure}[tp]
	\centering
	\includegraphics[width=.9\textwidth]{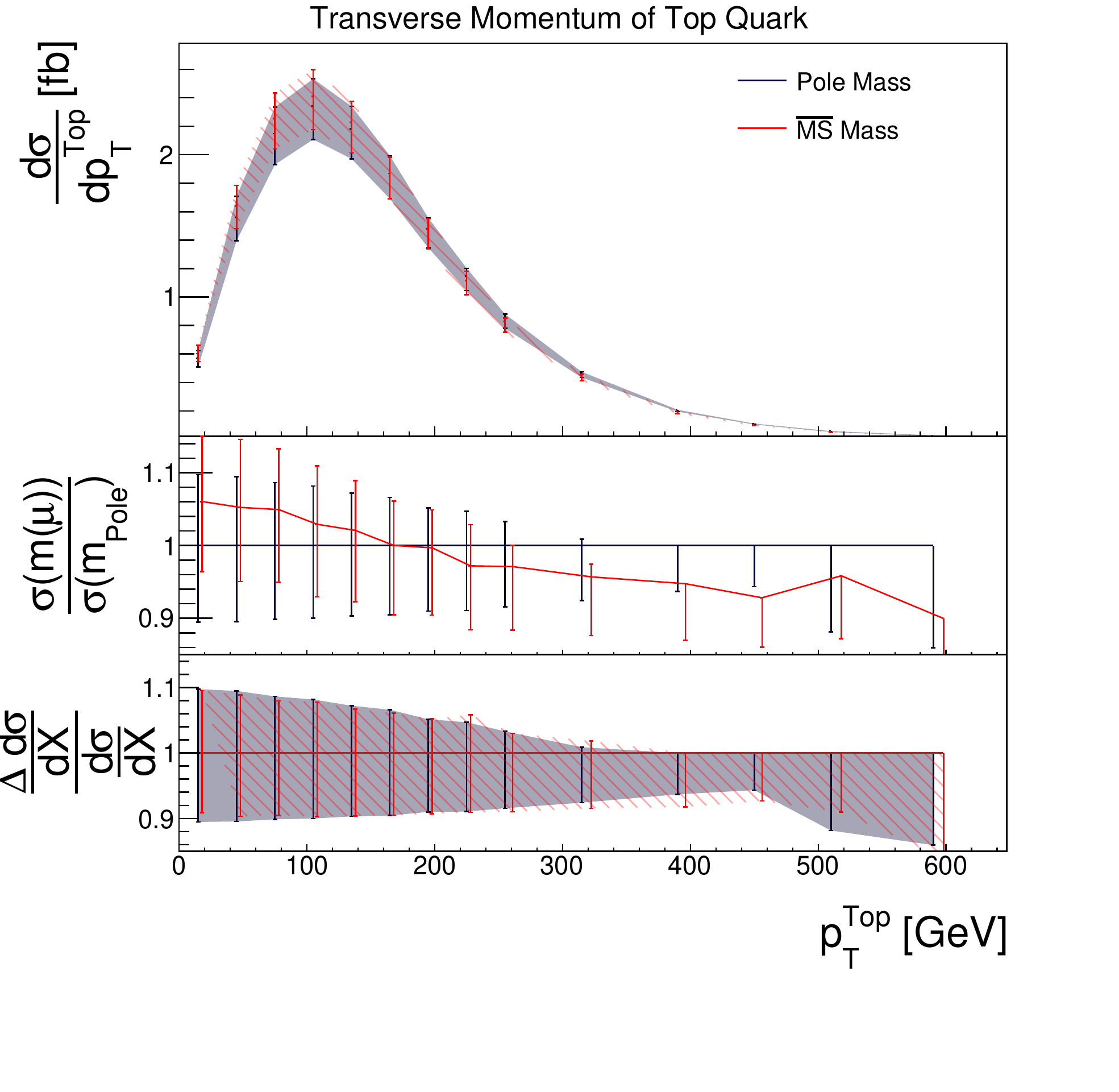}
	\caption{\label{fig:pttopSyst} Transverse momentum of the top quark in the \ttbarH\ process in the pole mass (blue) and the \msbar\ (red) schemes. 
		The ratio plot in the middle panel shows the relative shape difference between the two predictions normalized to the cross-section in the pole mass scheme. 
		In the lower panel, the relative uncertainty of the predictions normalized to the their individual mean values is shown. 
		The scale uncertainties are illustrated by the solid, dark blue and the hatched, red bands respectively.
		The uncertainty on the predictions is derived through the variation of \muR\ and \muF\ in the interval
		$[0.5\mu_0,\, 2\mu_0]$. 
		In the case of \msbar\ scheme, the statistical uncertainty on the calculation of the Born derivative is added in quadrature. 
		The cross-sections are calculated for the same $\pT^\text{Top}$ values in each bin for both distributions, but the bin centers in the ratio plots are shifted with respect to each other for better readability.}
\end{figure}

In Fig.~\ref{fig:pttopSyst} we illustrate the differential distributions 
of the \pT\ of the top quark in the \ttbarH\ production.
A difference in the shapes of the differential cross-sections in the pole mass and \msbar\ scheme is observed.
Compared to the differential cross-section in the pole mass scheme, the \msbar\ cross-section is shifted towards lower \pT\ values and shows a more pronounced (higher) peak.
Except for the bins in the range $400-450$ \GeV, all differences between the distributions are covered by the scale uncertainties. 
The differential cross-section with the top quark running mass 
show more stability over a wide range of \pT\ and reduce the scale uncertainties 
in the peak region slightly, where the bulk of the events is found.
However, the pole mass prediction appears to be slightly less sensitive to scale variations in the range $250-500$ \GeV.
This is related to the scale choice of $\muR=2m_{\text{t}}+m_{\text{H}}$ in the pole mass simulation.

\begin{figure}[tp]
	\centering
	\includegraphics[width=.9\textwidth]{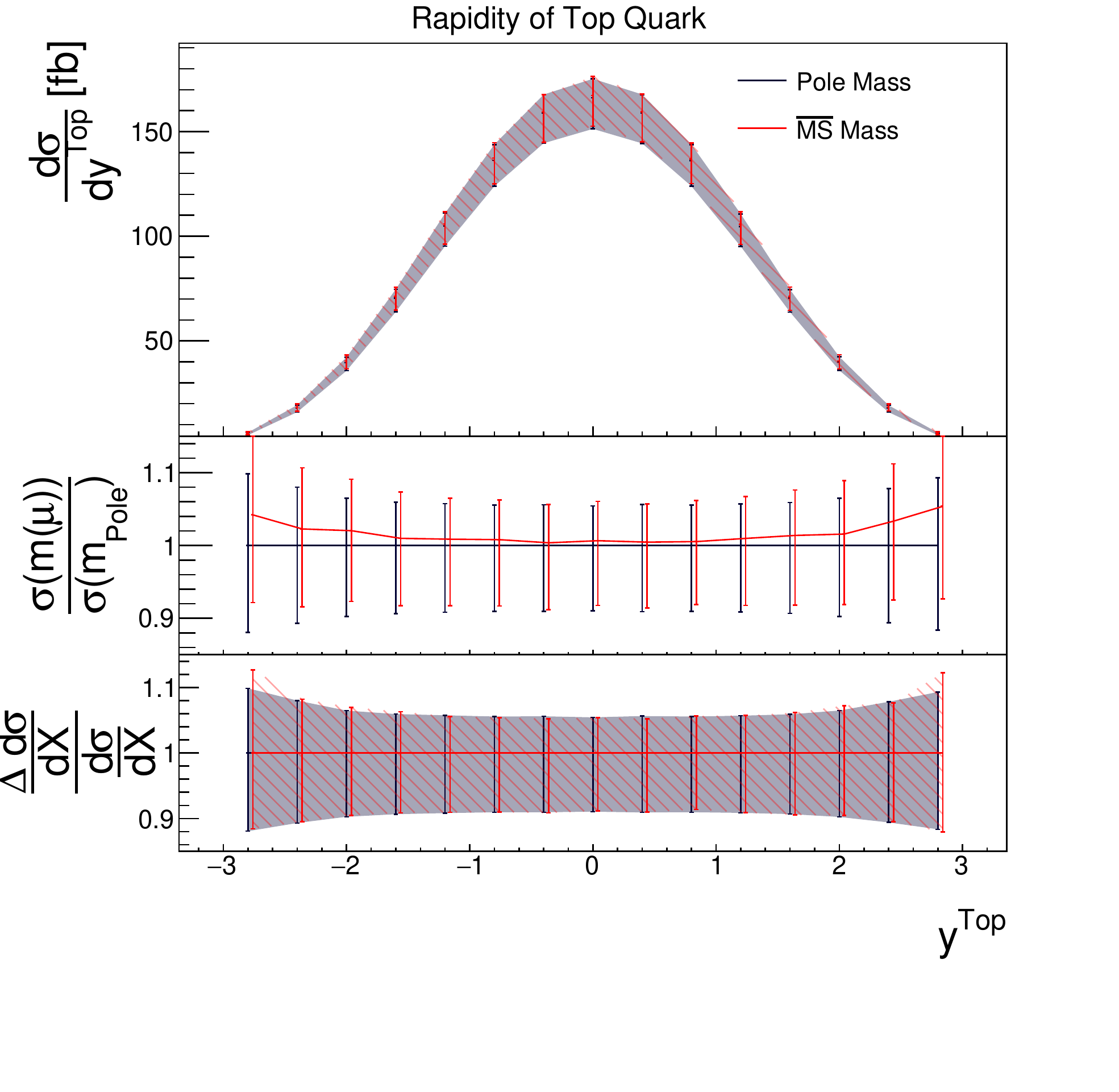}
	\caption{\label{fig:yttopSyst} 
	Same as Fig.~\ref{fig:pttopSyst} for rapidity of the top quark in the \ttbarH\ process in the pole mass (blue) and the \msbar\ (red) schemes.
	The cross-sections are calculated for the same $y^\text{Top}$ values in each bin for both distributions, but the 
	bin centers in the ratio plots are shifted with respect to each other for better readability.}
\end{figure}

The rapidity of the top quark in the \ttbarH\ production is illustrated in Fig.~\ref{fig:yttopSyst}.
Overall, the difference between the pole and the \msbar\ mass are not as pronounced in this distribution compared to the \pT\ of the top quarks.
Small differences in the shape of the distributions are observed.
Compared to the pole mass scheme, the \msbar\ calculation predicts a slightly higher yield of top quarks with large $|y^\text{Top}|$ values.
However, the differences of the calculated distributions are well in agreement with each other.
The relative uncertainties in both cases are similar in size across all rapidity ranges.

\begin{figure}[tp]
	\centering
	\includegraphics[width=.9\textwidth]{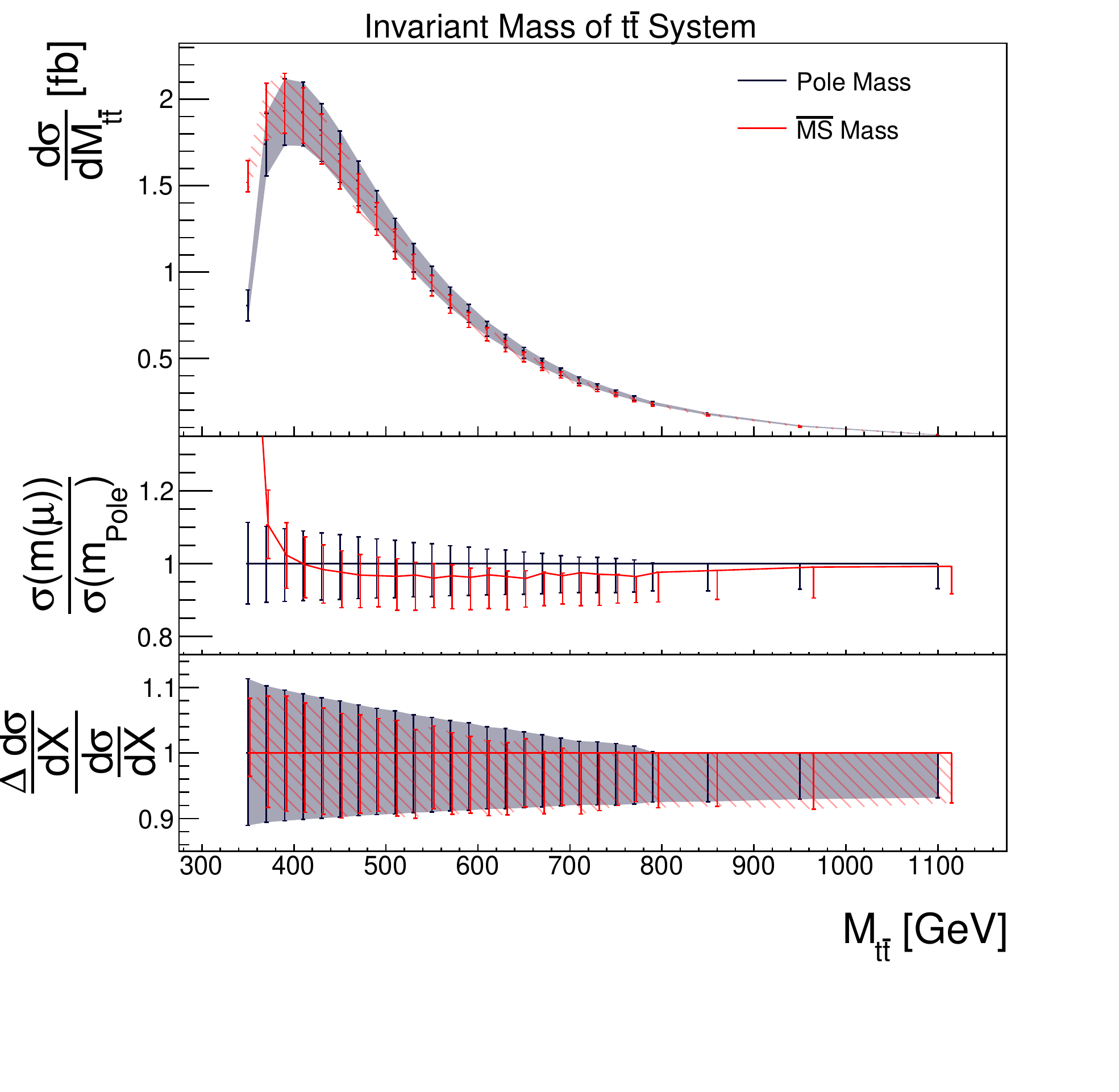}
	\caption{\label{fig:ttinvSyst} 
		Same as Fig.~\ref{fig:pttopSyst} for invariant mass of the \ttbar\ system in \ttbarH\ shown for the pole (blue) and \msbar\ mass (red) schemes. 
		The cross-sections are calculated for the same \Mttbar\ values in each bin for both distributions, 
	but the bin centers in the ratio plots are shifted with respect to each other for better readability.}
\end{figure}

In Fig.~\ref{fig:ttinvSyst} we display the differential cross-sections as a function of the invariant mass of the \ttbar\ system.
This distribution shows the largest deviations of all studied differential cross-sections between the pole and the running mass schemes.
The peak of the distribution in the \msbar\ scheme is shifted towards lower values of \Mttbar\ where differences between the calculations of more than 10\% are observed.
For $\Mttbar<800$ \GeV, the cross-sections with the top quark running is slightly less sensitive to scale variations.
The differences are mostly covered by the scale uncertainties, but the deviations in the threshold region (first bin) are significant.

The distributions in the kinematic variables of the top quark 
are especially interesting for studies of spin correlations 
in \ttbar~and \ttbar+X processes, where X denotes some additional observed final state, e.g. jet, boson etc.
Because of different couplings, dissimilar distributions 
of the top quark decay products are expected for the \ttbar+X processes, 
i.e., the radiation of a Higgs boson from a top quark in \ttbarH\ production flips the chirality of the top quark. 
This causes differences in the \pT\ distribution of the top quark in the \ttbarH\ process compared to \ttbar\ with additional radiation of a gluon~\cite{Biswas:2016wcb}.
In such analyses, it should be kept in mind that the chosen top quark mass renormalization scheme can have an impact on the shape of the 
distributions in both the \ttbar\ and \ttbarH\ processes.
Similar comments apply to the use of \ttbarH\ distributions when making statements about exclusion limits on parameters in models for physics beyond the SM.

\begin{figure}[tp]
	\centering
	\includegraphics[width=.9\textwidth]{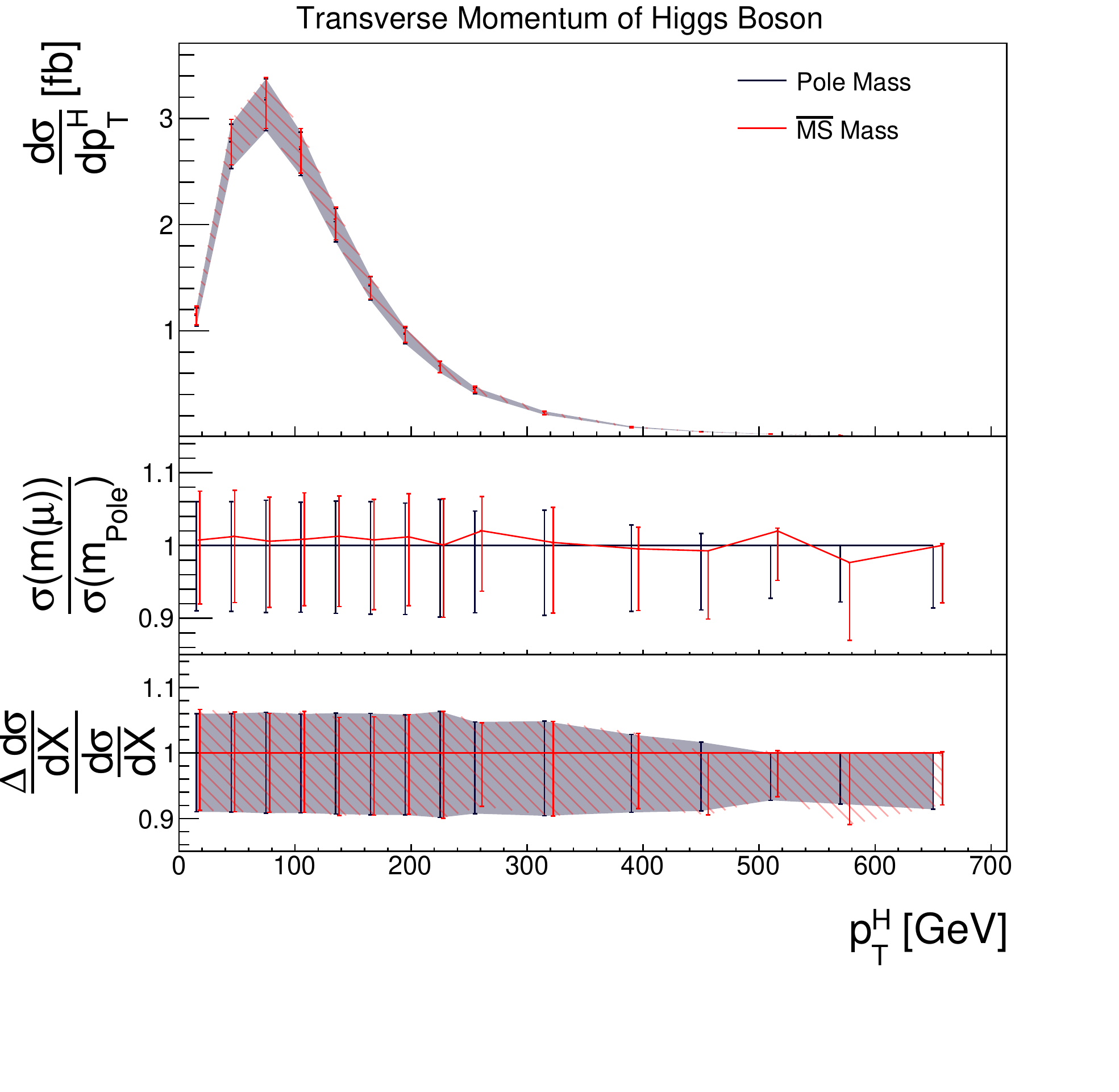}
	\caption{\label{fig:ptHSyst} Same as Fig.~\ref{fig:pttopSyst} for
	transverse momentum of the Higgs boson in \ttbarH\ production illustrated for the pole (blue) and \msbar\ mass (red) schemes. 
    The cross-sections are calculated for the same $\pT^\text{Higgs}$ values in each bin for both distributions, 
	but the bin centers in the ratio plots are shifted with respect to each other for better readability.}
\end{figure}

\begin{figure}[tp]
	\centering
	\includegraphics[width=.9\textwidth]{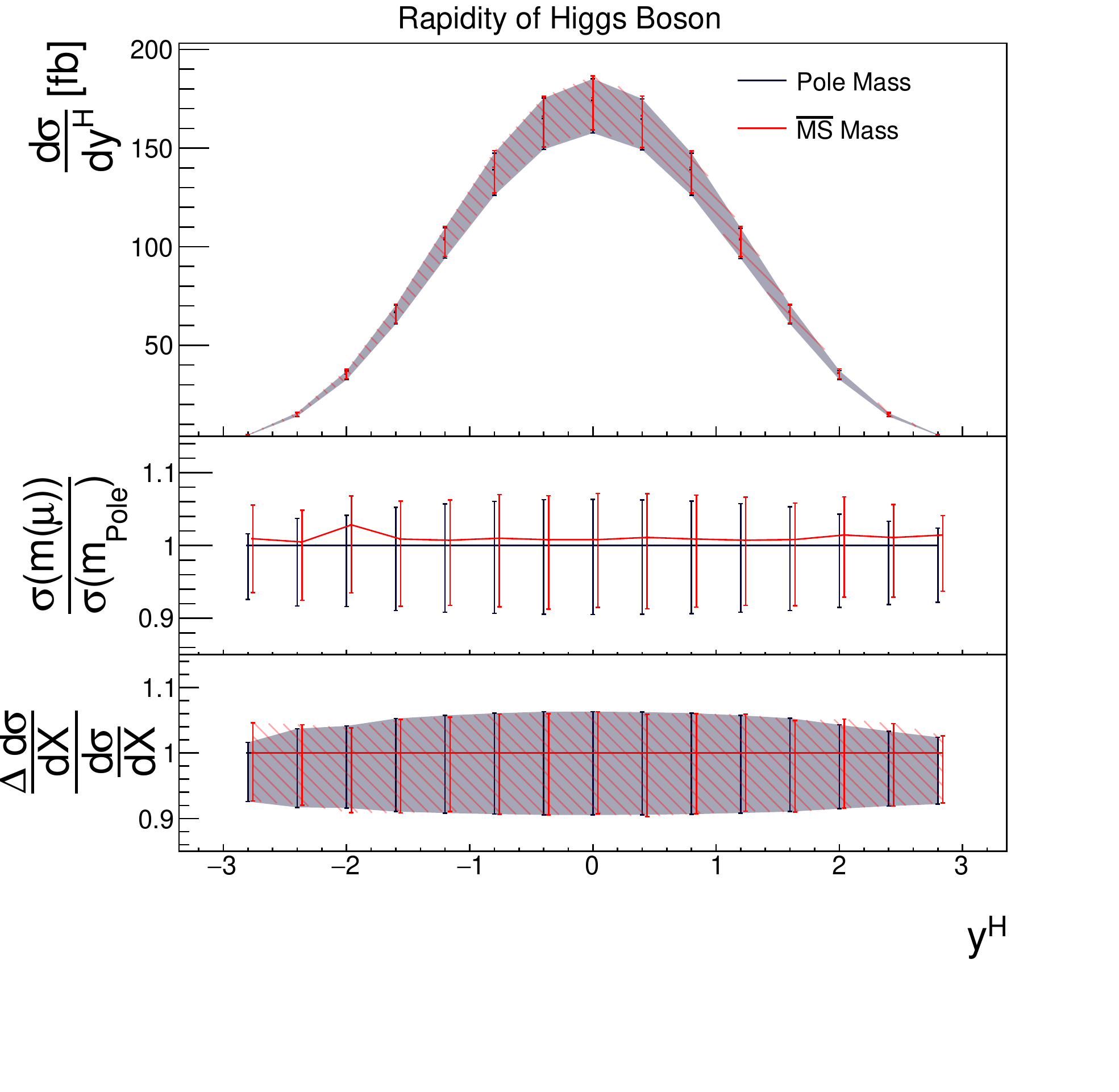}
	\caption{\label{fig:yHSyst} Same as Fig.~\ref{fig:pttopSyst} for
	rapidity of the Higgs boson in \ttbarH\ production illustrated for the pole (blue) and \msbar\ mass (red) schemes. 
    The cross-sections are calculated for the same $y^\text{H}$ values in each bin for both distributions, but the 
	bin centers in the ratio plots are shifted with respect to each other for better readability.}
\end{figure}

The behavior of the Higgs boson in \ttbarH\ production is studied 
in Figs.~\ref{fig:ptHSyst} and \ref{fig:yHSyst}, where we display distributions in the \pT\ and the rapidity of the Higgs boson.
A small systematic shift upwards ($\approx 1\%$) of the cross-sections with the \msbar\ scheme is observed in both cases.
The shape and the scale uncertainties on the cross-sections show negligible differences that can be attributed to statistical fluctuations.

The negligible impact of the top quark mass renormalization scheme, i.e. on-shell or \msbar, on the distributions related to the Higgs boson is expected, 
since the present study considers the QCD radiative corrections.
The electroweak corrections to \ttbarH\ production~\cite{Denner:2016wet}
and to the top quark mass renormalization should be incorporated into the studies for comparisons with the differential cross-sections from precision measurements in the future.

\begin{figure}[tp]
	\centering
	\includegraphics[width=.9\textwidth]{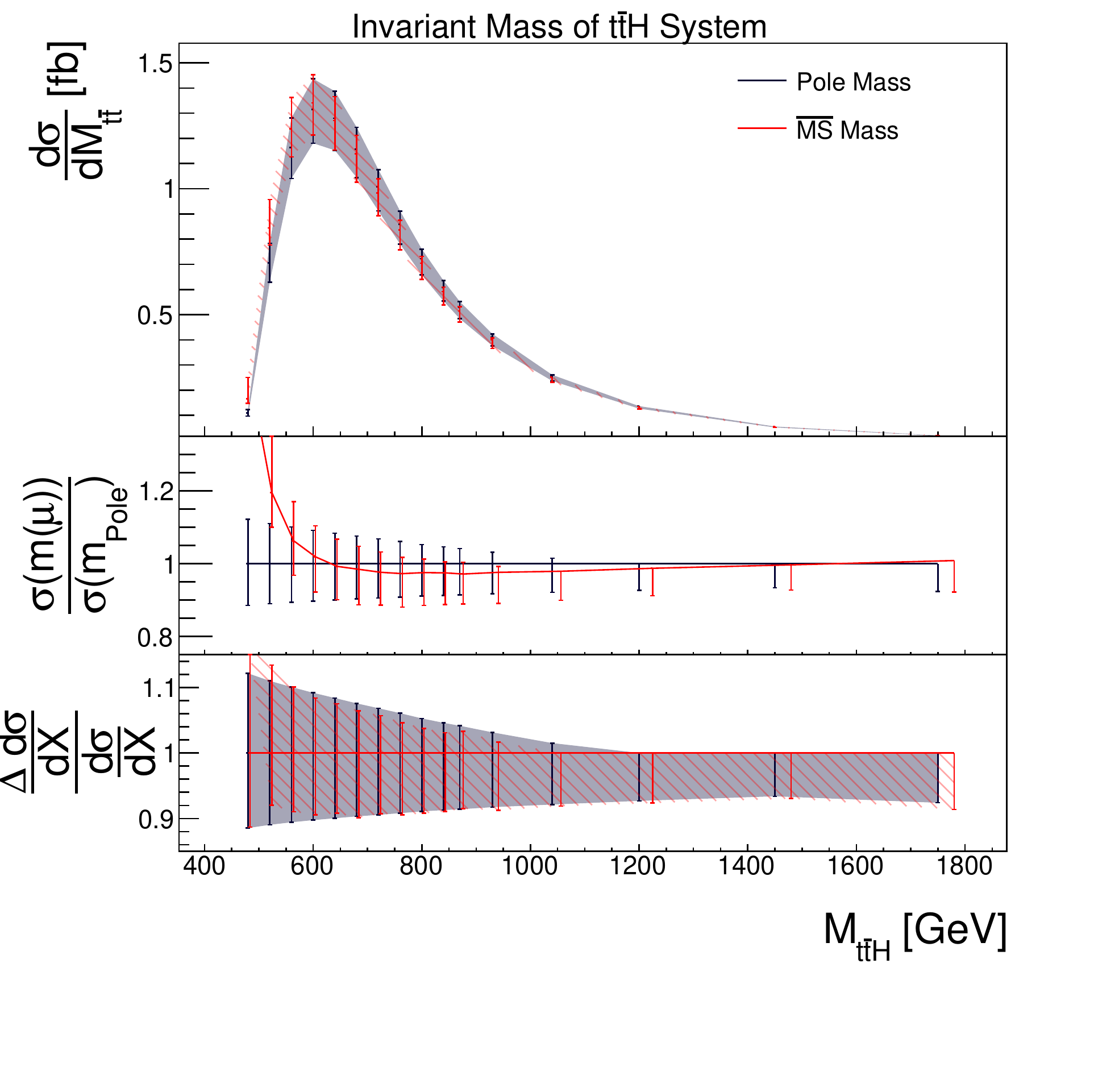}
	\caption{\label{fig:invmassSyst}  Same as Fig.~\ref{fig:pttopSyst} for invariant mass of the \ttbarH\ system shown for the pole (blue) and \msbar\ mass (red) schemes. 
	The cross-sections are calculated for the same $M_\text{\ttbarH}$ values in each bin for both distributions, 
	but the bin centers in the ratio plots are shifted with respect to each other for better readability.}
\end{figure}

In Fig. \ref{fig:invmassSyst}, 
the invariant mass of the \ttbarH\ system is presented.
The peak of the distribution is slightly higher and shifted towards lower values of \MttbarH, 
similar to what was observed for the \Mttbar\ distribution in Fig.~\ref{fig:ttinvSyst}.
The scale uncertainties in the \msbar\ scheme are marginally reduced in the range of $[600-1000]$ \GeV\ in \MttbarH.
At the production threshold ($[460-500]$\GeV, first bin), 
the scale uncertainties for the calculations in the \msbar\ scheme are significantly increased with the cross-section value in this bin 
being ${0.166}^{+51.14\%}_{-11.30\%}$ fb/(bin width) for the calculations in the \msbar\ mass 
and ${0.109}^{+12.16\%}_{-11.46\%}$ fb/(bin width) for the pole mass scheme.
This is an expected effect, as 
the \msbar\ mass displays shortcomings at the production threshold, 
where the calculations in the \msbar\ scheme are not well-behaved. 
These shortcomings of the \msbar\ mass have already been observed for the \ttbar\ production at lepton and hadron colliders~\cite{Hoang:2000yr,Dowling:2013baa}, 
and they are confirmed here also for the \ttbarH\ production process.
A possible solution for distributions close to threshold are the so-called \emph{low-scale short-distance masses}, such as the MSR mass.
The MSR mass~\cite{Hoang:2017suc,Butenschoen:2016lpz} was proposed for the usage in \ttbar\ cross-sections with the top quark running mass. 
It has the advantage that it is better behaved at the threshold and 
can be obtained from the \msbar\ mass in perturbative QCD similar to Eq.~(\ref{eq:MassRelation}), see also~\cite{Garzelli:2020fmd}.


\section{Summary}
\label{ch:PhenomenologyttH:sec:Summary}

The impact of the top quark \msbar\ mass on differential distributions in \ttbarH\ production is an interesting topic for future differential \ttbarH\ analyses at high precision, in particular at the HL-LHC.
We have presented the first ever studies of the impact of the 
top quark mass in the \msbar\ scheme on the \ttbarH\ production cross-sections.
Our work compares fixed-order NLO QCD predictions with stable top quarks in the \msbar\ and the pole mass scheme for differential cross-sections as a function of \pT\ and $y$ of the top quark and the Higgs boson, as well as the invariant mass of the \ttbar\ and the \ttbarH\ systems.

The overall impact of the top quark mass in the \msbar\ scheme on the \ttbarH\ production is found to be small.
The behavior of the Higgs boson is barely affected by the 
change of the quark mass renormalization scheme.
However, for the top quark differences in the shape of the \pT\ and $y$ distributions are visible in comparisons between the \msbar\ and the pole mass schemes. 
Additionally, the scale uncertainties in the \msbar\ scheme are slightly reduced in the low-\pT\ region of the top quark \pT\ distribution.
The largest corrections due to the effects of the top quark \msbar\ mass are observed in the invariant mass distributions of the \ttbar\ and the \ttbarH\ systems.
Over a wide range of the distributions, a moderate decrease in scale uncertainties for the \msbar\ scheme is seen.
However, also the expected shortcomings of the \msbar\ mass at the threshold of the \MttbarH\ distribution become apparent. 
In summary, it can be confirmed that no major systematic effects due to the choice of the quark mass renormalization scheme were neglected in the experimental analyses leading to the observation of the \ttbarH\ production in the year 2018. 

Overall, the numerical approach for the estimation of the Born-level mass derivative is advantageous since it enables also the calculation of other differential cross-sections within the same framework, e.g. \ttbar, \ttZ, \ttbarH, \ttgamma, etc., 
without requiring any analytical expressions for the derivative of the Born cross-sections. 
On the other hand, those calculations have to be performed for a range of values of the top quark mass at a high numerical precision and, therefore, the approach is computationally expensive.
For this reason, the approach can be improved by incorporating the analytical expressions for the mass derivative of the Born cross-sections.

The present studies focus on the NLO QCD radiative corrections. Further improvements of the cross-section predictions 
due to the effect of resummed threshold logarithms, Coulomb corrections or the complete NNLO QCD corrections 
will become important as the precision of differential measurements for \ttbarH\ production increases.
The studies can also be improved by incorporating the electroweak corrections to \ttbarH\ production and the 
renormalization of the top quark mass in the future.
The shortcomings of the \msbar\ mass at the production threshold can be overcome by considering a low-scale short-distance mass, such as the MSR mass.
We leave theses aspects for future studies.

\if{1=0}
At the size of the currently available data sets recorded by the ATLAS and CMS collaborations ($\approx 137\invfb$) during the \thirteenTeV\ runs of the Large Hadron Collider, no precise differential \ttbarH\ measurements can be expected.
The uncertainties on the backgrounds in the current data analyses of \ttbarH\ production outweigh the uncertainties on the signal modeling by far.
The \ttbarH\ analysis in the \Hbb\ decay channel (\ttbarHbb) was the most sensitive to the signal strength of the \ttbarH\ production process in the combination of the CMS collaboration that led to the observation of the \ttbarH\ production and, therefore, to the observation of the top-Higgs Yukawa coupling.
The results of the \ttbarHbb\ analysis depend to a large amount on the uncertainties related to the identification and modeling of the top quark pair production with additional jets originating from the hadronization of \ab-quarks (\ttbjets) \cite{Sirunyan:2018mvw}.
\fi

\acknowledgments
S.M. acknowledges support from Bundesministerium f\"ur Bildung und Forschung (contract 05H21GUCCA).

\appendix
\section{Running and Pole Mass Cross-Sections}
\label{app:RunningCrossSection}
In this Section, the differential cross-sections for \ttbarH\ production calculated in the \msbar\ mass scheme at NLO in QCD are displayed. 
The \msbar\ scheme values are compared with those in the pole mass scheme in the corresponding tables.
Tables~\ref{tab:comparison_RunningMass_ptTop}-\ref{tab:comparison_RunningMass_MttH} contain the differential cross-sections in each bin 
divided by the bin width and, thus, the values correspond to the Figures in Section \ref{ch:PhenomenologyttH:sec:Results}.

\begin{table}[hbt] 
	\centering
	\caption{\small Comparisons of differential cross-sections of \ttbarH\ production as a function of the \emph{\pT\ of the top quark}  calculated in the pole and the \msbar\ mass schemes.
	The cross-section values are divided by the width of the corresponding bin.}
	\label{tab:comparison_RunningMass_ptTop}
	\begin{tabular}{lcc}\toprule
			& \multicolumn{2}{c}{Cross-Section [fb/(bin width)]} 
			\\\cmidrule(lr){2-3}
			Bin range [\GeV]	& 	Pole Mass 										& \msbar\ Mass 				\\\midrule
			$[0.00,30.00]$ 	 & 	  ${0.568}^{+0.055(+9.73\%)}_{-0.060(-10.53\%)}$ 	& 	 ${0.615}^{+0.050(+8.09\%)}_{-0.051(-8.33\%)}$  \\ 
			$[30.00,60.00]$ 	 & 	  ${1.559}^{+0.148(+9.48\%)}_{-0.163(-10.46\%)}$ & 	 ${1.670}^{+0.127(+7.63\%)}_{-0.149(-8.95\%)}$  \\ 
			$[60.00,90.00]$ 	 & 	  ${2.148}^{+0.186(+8.64\%)}_{-0.218(-10.13\%)}$ & 	 ${2.286}^{+0.161(+7.04\%)}_{-0.206(-8.99\%)}$  \\ 
			$[90.00,120.00]$ 	 & 	  ${2.342}^{+0.191(+8.16\%)}_{-0.234(-10.01\%)}$ & 	 ${2.434}^{+0.174(+7.16\%)}_{-0.229(-9.40\%)}$  \\ 
			$[120.00,150.00]$ 	 & 	  ${2.182}^{+0.157(+7.21\%)}_{-0.211(-9.67\%)}$ & 	 ${2.238}^{+0.144(+6.44\%)}_{-0.215(-9.61\%)}$  \\ 
			$[150.00,180.00]$ 	 & 	  ${1.868}^{+0.123(+6.60\%)}_{-0.178(-9.54\%)}$ & 	 ${1.865}^{+0.116(+6.24\%)}_{-0.179(-9.60\%)}$  \\ 
			$[180.00,210.00]$ 	 & 	  ${1.479}^{+0.076(+5.14\%)}_{-0.133(-9.00\%)}$ & 	 ${1.463}^{+0.085(+5.81\%)}_{-0.141(-9.65\%)}$  \\ 
			$[210.00,240.00]$ 	 & 	  ${1.149}^{+0.054(+4.69\%)}_{-0.103(-8.94\%)}$ & 	 ${1.103}^{+0.073(+6.60\%)}_{-0.106(-9.57\%)}$  \\ 
			$[240.00,270.00]$ 	 & 	  ${0.852}^{+0.028(+3.30\%)}_{-0.072(-8.44\%)}$ & 	 ${0.814}^{+0.033(+4.02\%)}_{-0.078(-9.53\%)}$  \\ 
			$[270.00,360.00]$ 	 & 	  ${0.470}^{+0.004(+0.88\%)}_{-0.036(-7.58\%)}$ & 	 ${0.438}^{+0.015(+3.42\%)}_{-0.041(-9.29\%)}$  \\ 
			$[360.00,420.00]$ 	 & 	  ${0.208}^{+0.000(+0.00\%)}_{-0.013(-6.31\%)}$ & 	 ${0.190}^{+0.000(+0.00\%)}_{-0.017(-9.18\%)}$  \\ 
			$[420.00,480.00]$ 	 & 	  ${0.111}^{+0.000(+0.00\%)}_{-0.006(-5.67\%)}$ & 	 ${0.099}^{+0.000(+0.00\%)}_{-0.007(-7.55\%)}$  \\ 
			$[480.00,540.00]$ 	 & 	  ${0.058}^{+0.000(+0.00\%)}_{-0.007(-11.84\%)}$ & 	 ${0.053}^{+0.000(+0.00\%)}_{-0.003(-6.49\%)}$  \\ 
			$[540.00,640.00]$ 	 & 	  ${0.029}^{+0.000(+0.00\%)}_{-0.004(-14.07\%)}$ & 	 ${0.024}^{+0.000(+0.00\%)}_{-0.003(-14.11\%)}$  \\ \bottomrule
				\end{tabular}
\end{table}

\begin{table}[hbt] 
	\centering
	\caption{\small 
	Same as Tab.~\ref{tab:comparison_RunningMass_ptTop} for the \emph{rapidity of the top quark}.
%
}
	\label{tab:comparison_RunningMass_yTop}
	\begin{tabular}{lcc}\toprule
		& \multicolumn{2}{c}{Cross-Section [fb/(bin width)]} 
		\\\cmidrule(lr){2-3}
		Bin range 	& 	Pole Mass 										& \msbar\ Mass 				\\\midrule
		 $[-3.00,-2.60]$ 	 & 	  ${5.649}^{+0.557(+9.87\%)}_{-0.674(-11.94\%)}$ & 	 ${5.886}^{+0.747(+12.69\%)}_{-0.681(-11.57\%)}$  \\ 
		$[-2.60,-2.20]$ 	 & 	  ${17.925}^{+1.434(+8.00\%)}_{-1.916(-10.69\%)}$ & 	 ${18.332}^{+1.506(+8.21\%)}_{-1.919(-10.47\%)}$  \\ 
		$[-2.20, -1.80]$ 	 & 	  ${39.628}^{+2.563(+6.47\%)}_{-3.858(-9.74\%)}$ & 	 ${40.433}^{+2.808(+6.94\%)}_{-3.848(-9.52\%)}$  \\ 
		$[-1.80, -1.40]$ 	 & 	  ${70.473}^{+4.187(+5.94\%)}_{-6.586(-9.35\%)}$ & 	 ${71.164}^{+4.472(+6.28\%)}_{-6.507(-9.14\%)}$  \\ 
		$[-1.40, -1.00]$ 	 & 	  ${104.944}^{+6.036(+5.75\%)}_{-9.630(-9.18\%)}$ & 	 ${105.851}^{+5.885(+5.56\%)}_{-9.604(-9.07\%)}$  \\ 
		$[-1.00, -0.60]$ 	 & 	  ${136.231}^{+7.580(+5.56\%)}_{-12.327(-9.05\%)}$ & 	 ${137.323}^{+7.419(+5.40\%)}_{-12.399(-9.03\%)}$  \\ 
		$[-0.60, -0.20]$ 	 & 	  ${158.782}^{+8.893(+5.60\%)}_{-14.365(-9.05\%)}$ & 	 ${159.370}^{+8.339(+5.23\%)}_{-14.596(-9.16\%)}$  \\ 
		$[-0.20,0.20]$ 	 & 	  ${166.173}^{+9.038(+5.44\%)}_{-14.890(-8.96\%)}$ & 	 ${167.266}^{+8.995(+5.38\%)}_{-14.743(-8.81\%)}$  \\ 
		$[0.20,0.60]$ 	 & 	  ${158.765}^{+8.950(+5.64\%)}_{-14.405(-9.07\%)}$ & 	 ${159.494}^{+8.353(+5.24\%)}_{-14.373(-9.01\%)}$  \\ 
		$[0.60,1.00]$ 	 & 	  ${136.198}^{+7.579(+5.56\%)}_{-12.315(-9.04\%)}$ & 	 ${136.928}^{+7.710(+5.63\%)}_{-11.811(-8.63\%)}$  \\ 
		$[1.00,1.40]$ 	 & 	  ${104.680}^{+5.956(+5.69\%)}_{-9.569(-9.14\%)}$ & 	 ${105.699}^{+6.041(+5.72\%)}_{-9.624(-9.10\%)}$  \\ 
		$[1.40,1.80]$ 	 & 	  ${70.341}^{+4.149(+5.90\%)}_{-6.558(-9.32\%)}$ & 	 ${71.311}^{+4.386(+6.15\%)}_{-6.746(-9.46\%)}$  \\ 
		$[1.80,2.20]$ 	 & 	  ${39.825}^{+2.579(+6.48\%)}_{-3.888(-9.76\%)}$ & 	 ${40.449}^{+2.923(+7.23\%)}_{-3.859(-9.54\%)}$  \\ 
		$[2.20,2.60]$ 	 & 	  ${17.856}^{+1.394(+7.81\%)}_{-1.893(-10.60\%)}$ & 	 ${18.452}^{+1.406(+7.62\%)}_{-1.935(-10.49\%)}$  \\ 
		$[2.60,3.00]$ 	 & 	  ${5.575}^{+0.519(+9.30\%)}_{-0.649(-11.64\%)}$ & 	 ${5.874}^{+0.720(+12.26\%)}_{-0.709(-12.06\%)}$  \\ \bottomrule
	\end{tabular}
\end{table}

\begin{table}[hbt] 
	\centering
	\caption{\small 
Same as Tab.~\ref{tab:comparison_RunningMass_ptTop} for the
\emph{invariant mass \Mttbar\ of the \ttbar\ system}.
}
	\label{tab:comparison_RunningMass_Mttbar}
	\begin{tabular}{lcc}\toprule
		& \multicolumn{2}{c}{Cross-Section [fb/(bin width)]} 
		\\\cmidrule(lr){2-3}
		Bin range [\GeV]	& 	Pole Mass 										& \msbar\ Mass 				\\\midrule
		$[460.00,480.00]$ 	 & 	  ${1.530}^{+0.112(+7.35\%)}_{-0.147(-9.57\%)}$ & 	 ${1.482}^{+0.086(+5.83\%)}_{-0.136(-9.21\%)}$  \\ 
		$[480.00,500.00]$ 	 & 	  ${1.377}^{+0.094(+6.82\%)}_{-0.130(-9.42\%)}$ & 	 ${1.331}^{+0.070(+5.27\%)}_{-0.118(-8.89\%)}$  \\ 
		$[500.00,520.00]$ 	 & 	  ${1.233}^{+0.079(+6.42\%)}_{-0.115(-9.31\%)}$ & 	 ${1.190}^{+0.059(+4.99\%)}_{-0.115(-9.64\%)}$  \\ 
		$[520.00,540.00]$ 	 & 	  ${1.101}^{+0.064(+5.84\%)}_{-0.100(-9.11\%)}$ & 	 ${1.066}^{+0.038(+3.58\%)}_{-0.106(-9.95\%)}$  \\ 
		$[540.00,560.00]$ 	 & 	  ${0.981}^{+0.053(+5.44\%)}_{-0.088(-9.00\%)}$ & 	 ${0.942}^{+0.039(+4.13\%)}_{-0.080(-8.47\%)}$  \\ 
		$[560.00,580.00]$ 	 & 	  ${0.870}^{+0.043(+4.93\%)}_{-0.077(-8.82\%)}$ & 	 ${0.841}^{+0.026(+3.10\%)}_{-0.079(-9.37\%)}$  \\ 
		$[580.00,600.00]$ 	 & 	  ${0.777}^{+0.036(+4.58\%)}_{-0.068(-8.75\%)}$ & 	 ${0.748}^{+0.019(+2.54\%)}_{-0.069(-9.21\%)}$  \\ 
		$[600.00,620.00]$ 	 & 	  ${0.687}^{+0.027(+3.98\%)}_{-0.059(-8.54\%)}$ & 	 ${0.666}^{+0.013(+1.88\%)}_{-0.064(-9.57\%)}$  \\ 
		$[620.00,640.00]$ 	 & 	  ${0.616}^{+0.023(+3.73\%)}_{-0.052(-8.51\%)}$ & 	 ${0.594}^{+0.009(+1.60\%)}_{-0.056(-9.44\%)}$  \\ 
		$[640.00,660.00]$ 	 & 	  ${0.546}^{+0.018(+3.24\%)}_{-0.046(-8.37\%)}$ & 	 ${0.524}^{+0.011(+2.18\%)}_{-0.043(-8.23\%)}$  \\ 
		$[660.00,680.00]$ 	 & 	  ${0.486}^{+0.014(+2.79\%)}_{-0.040(-8.23\%)}$ & 	 ${0.474}^{+0.001(+0.16\%)}_{-0.044(-9.28\%)}$  \\ 
		$[680.00,700.00]$ 	 & 	  ${0.434}^{+0.010(+2.25\%)}_{-0.035(-8.05\%)}$ & 	 ${0.420}^{+0.003(+0.75\%)}_{-0.034(-7.99\%)}$  \\ 
		$[700.00,720.00]$ 	 & 	  ${0.386}^{+0.007(+1.80\%)}_{-0.031(-7.91\%)}$ & 	 ${0.377}^{+0.000(+0.00\%)}_{-0.035(-9.31\%)}$  \\ 
		$[720.00,740.00]$ 	 & 	  ${0.348}^{+0.006(+1.71\%)}_{-0.028(-7.94\%)}$ & 	 ${0.338}^{+0.000(+0.00\%)}_{-0.030(-8.80\%)}$  \\ 
		$[740.00,760.00]$ 	 & 	  ${0.313}^{+0.005(+1.44\%)}_{-0.025(-7.95\%)}$ & 	 ${0.303}^{+0.000(+0.10\%)}_{-0.024(-8.02\%)}$  \\ 
		$[760.00,780.00]$ 	 & 	  ${0.279}^{+0.003(+1.03\%)}_{-0.022(-7.79\%)}$ & 	 ${0.269}^{+0.000(+0.00\%)}_{-0.020(-7.42\%)}$  \\ 
		$[780.00,800.00]$ 	 & 	  ${0.250}^{+0.000(+0.16\%)}_{-0.019(-7.49\%)}$ & 	 ${0.244}^{+0.000(+0.00\%)}_{-0.020(-8.35\%)}$  \\ 
		$[800.00,900.00]$ 	 & 	  ${0.185}^{+0.000(+0.00\%)}_{-0.014(-7.43\%)}$ & 	 ${0.182}^{+0.000(+0.00\%)}_{-0.015(-8.12\%)}$  \\ 
		$[900.00,1000.00]$ 	 & 	  ${0.113}^{+0.000(+0.00\%)}_{-0.008(-7.05\%)}$ & 	 ${0.112}^{+0.000(+0.00\%)}_{-0.010(-8.56\%)}$  \\ 
		$[1000.00,1200.00]$ 	 & 	  ${0.059}^{+0.000(+0.00\%)}_{-0.004(-6.83\%)}$ & 	 ${0.058}^{+0.000(+0.00\%)}_{-0.004(-7.62\%)}$  \\ \bottomrule
	\end{tabular}
\end{table}

\begin{table}[hbt] 
	\centering
	\caption{\small 
Same as Tab.~\ref{tab:comparison_RunningMass_ptTop} for the	
\emph{\pT\ of the Higgs boson}.
}
	\label{tab:comparison_RunningMass_ptH}
	\begin{tabular}{lcc}\toprule
		& \multicolumn{2}{c}{Cross-Section [fb/(bin width)]} 
		\\\cmidrule(lr){2-3}
		Bin range [\GeV]	& 	Pole Mass 										& \msbar\ Mass 				\\\midrule
		 $[0.00,30.00]$ 	 & 	  ${1.148}^{+0.069(+6.00\%)}_{-0.103(-8.98\%)}$ & 	 ${1.157}^{+0.077(+6.66\%)}_{-0.101(-8.75\%)}$  \\ 
		$[30.00,60.00]$ 	 & 	  ${2.780}^{+0.167(+6.00\%)}_{-0.251(-9.03\%)}$ & 	 ${2.815}^{+0.177(+6.28\%)}_{-0.253(-8.97\%)}$  \\ 
		$[60.00,90.00]$ 	 & 	  ${3.176}^{+0.197(+6.21\%)}_{-0.292(-9.20\%)}$ & 	 ${3.194}^{+0.192(+6.03\%)}_{-0.289(-9.05\%)}$  \\ 
		$[90.00,120.00]$ 	 & 	  ${2.709}^{+0.161(+5.95\%)}_{-0.248(-9.16\%)}$ & 	 ${2.733}^{+0.173(+6.33\%)}_{-0.248(-9.06\%)}$  \\ 
		$[120.00,150.00]$ 	 & 	  ${2.027}^{+0.123(+6.08\%)}_{-0.189(-9.32\%)}$ & 	 ${2.053}^{+0.112(+5.43\%)}_{-0.196(-9.54\%)}$  \\ 
		$[150.00,180.00]$ 	 & 	  ${1.424}^{+0.086(+6.03\%)}_{-0.134(-9.43\%)}$ & 	 ${1.435}^{+0.079(+5.51\%)}_{-0.136(-9.49\%)}$  \\ 
		$[180.00,210.00]$ 	 & 	  ${0.971}^{+0.057(+5.84\%)}_{-0.092(-9.47\%)}$ & 	 ${0.982}^{+0.057(+5.85\%)}_{-0.092(-9.34\%)}$  \\ 
		$[210.00,240.00]$ 	 & 	  ${0.671}^{+0.043(+6.34\%)}_{-0.066(-9.85\%)}$ & 	 ${0.671}^{+0.043(+6.35\%)}_{-0.067(-9.94\%)}$  \\ 
		$[240.00,270.00]$ 	 & 	  ${0.447}^{+0.021(+4.75\%)}_{-0.041(-9.25\%)}$ & 	 ${0.456}^{+0.021(+4.60\%)}_{-0.037(-8.13\%)}$  \\ 
		$[270.00,360.00]$ 	 & 	  ${0.231}^{+0.011(+4.85\%)}_{-0.022(-9.59\%)}$ & 	 ${0.232}^{+0.011(+4.82\%)}_{-0.022(-9.64\%)}$  \\ 
		$[360.00,420.00]$ 	 & 	  ${0.096}^{+0.003(+2.81\%)}_{-0.009(-9.07\%)}$ & 	 ${0.096}^{+0.003(+3.00\%)}_{-0.008(-8.51\%)}$  \\ 
		$[420.00,480.00]$ 	 & 	  ${0.052}^{+0.001(+1.67\%)}_{-0.005(-8.86\%)}$ & 	 ${0.051}^{+0.000(+0.00\%)}_{-0.005(-9.43\%)}$  \\ 
		$[480.00,540.00]$ 	 & 	  ${0.027}^{+0.000(+0.00\%)}_{-0.002(-7.23\%)}$ & 	 ${0.027}^{+0.000(+0.38\%)}_{-0.002(-6.66\%)}$  \\ 
		$[540.00,600.00]$ 	 & 	  ${0.016}^{+0.000(+0.00\%)}_{-0.001(-7.78\%)}$ & 	 ${0.016}^{+0.000(+0.00\%)}_{-0.002(-10.94\%)}$  \\ 
		$[600.00,700.00]$ 	 & 	  ${0.007}^{+0.000(+0.00\%)}_{-0.001(-8.60\%)}$ & 	 ${0.007}^{+0.000(+0.20\%)}_{-0.001(-7.92\%)}$  \\  \bottomrule
	\end{tabular}
\end{table}

\begin{table}[hbt] 
	\centering
	\caption{\small 
Same as Tab.~\ref{tab:comparison_RunningMass_ptTop} for the
\emph{rapidity of the Higgs boson}.
}
	\label{tab:comparison_RunningMass_yH}
	\begin{tabular}{lcc}\toprule
		& \multicolumn{2}{c}{Cross-Section [fb/(bin width)]} 
		\\\cmidrule(lr){2-3}
		Bin range 	& 	Pole Mass 										& \msbar\ Mass 				\\\midrule
		$[-3.00,-2.60]$ 	 & 	  ${4.636}^{+0.075(+1.61\%)}_{-0.344(-7.42\%)}$ & 	 ${4.679}^{+0.215(+4.59\%)}_{-0.342(-7.31\%)}$  \\ 
		$[-2.60,-2.20]$ 	 & 	  ${15.269}^{+0.567(+3.71\%)}_{-1.266(-8.29\%)}$ & 	 ${15.343}^{+0.663(+4.32\%)}_{-1.224(-7.98\%)}$  \\ 
		$[-2.20,-1.80]$ 	 & 	  ${35.476}^{+1.464(+4.13\%)}_{-2.978(-8.40\%)}$ & 	 ${36.482}^{+1.402(+3.84\%)}_{-3.319(-9.10\%)}$  \\ 
		$[-1.80,-1.40]$ 	 & 	  ${66.781}^{+3.510(+5.26\%)}_{-5.974(-8.95\%)}$ & 	 ${67.368}^{+3.472(+5.15\%)}_{-6.170(-9.16\%)}$  \\ 
		$[-1.40,-1.00]$ 	 & 	  ${103.670}^{+5.933(+5.72\%)}_{-9.498(-9.16\%)}$ & 	 ${104.423}^{+5.739(+5.50\%)}_{-9.298(-8.90\%)}$  \\ 
		$[-1.00,-0.60]$ 	 & 	  ${139.005}^{+8.423(+6.06\%)}_{-12.966(-9.33\%)}$ & 	 ${140.393}^{+8.357(+5.95\%)}_{-13.110(-9.34\%)}$  \\ 
		$[-0.60,-0.20]$ 	 & 	  ${164.915}^{+10.375(+6.29\%)}_{-15.588(-9.45\%)}$ & 	 ${166.222}^{+9.998(+6.01\%)}_{-15.688(-9.44\%)}$  \\ 
		$[-0.20,0.20]$ 	 & 	  ${174.130}^{+10.995(+6.31\%)}_{-16.493(-9.47\%)}$ & 	 ${175.535}^{+11.049(+6.29\%)}_{-16.237(-9.25\%)}$  \\ 
		$[0.20,0.60]$ 	 & 	  ${164.665}^{+10.275(+6.24\%)}_{-15.520(-9.43\%)}$ & 	 ${166.492}^{+9.871(+5.93\%)}_{-16.154(-9.70\%)}$  \\ 
		$[0.60,1.00]$ 	 & 	  ${138.981}^{+8.456(+6.08\%)}_{-12.980(-9.34\%)}$ & 	 ${140.206}^{+8.394(+5.99\%)}_{-12.986(-9.26\%)}$  \\ 
		$[1.00,1.40]$ 	 & 	  ${103.468}^{+5.935(+5.74\%)}_{-9.482(-9.16\%)}$ & 	 ${104.218}^{+6.136(+5.89\%)}_{-9.266(-8.89\%)}$  \\ 
		$[1.40,1.80]$ 	 & 	  ${66.802}^{+3.542(+5.30\%)}_{-5.977(-8.95\%)}$ & 	 ${67.349}^{+3.344(+4.96\%)}_{-6.065(-9.01\%)}$  \\ 
		$[1.80,2.20]$ 	 & 	  ${35.673}^{+1.537(+4.31\%)}_{-3.036(-8.51\%)}$ & 	 ${36.188}^{+1.867(+5.16\%)}_{-3.046(-8.42\%)}$  \\ 
		$[2.20,2.60]$ 	 & 	  ${15.172}^{+0.504(+3.32\%)}_{-1.230(-8.10\%)}$ & 	 ${15.338}^{+0.686(+4.47\%)}_{-1.241(-8.09\%)}$  \\ 
		$[2.60,3.00]$ 	 & 	  ${4.690}^{+0.112(+2.40\%)}_{-0.366(-7.81\%)}$ & 	 ${4.758}^{+0.125(+2.63\%)}_{-0.362(-7.61\%)}$  \\  \bottomrule
	\end{tabular}
\end{table}

\begin{table}[hbt] 
	\centering
	\caption{\small 
Same as Tab.~\ref{tab:comparison_RunningMass_ptTop} for the
\emph{invariant mass \MttbarH\ of the \ttbarH\ system}.
}
	\label{tab:comparison_RunningMass_MttH}
	\begin{tabular}{lcc}\toprule
		& \multicolumn{2}{c}{Cross-Section [fb/(bin width)]} 
		\\\cmidrule(lr){2-3}
		Bin range [\GeV]	& 	Pole Mass 										& \msbar\ Mass 				\\\midrule
		$[460.00,500.00]$ 	 & 	  ${0.109}^{+0.013(+12.16\%)}_{-0.013(-11.46\%)}$ & 	 ${0.166}^{+0.085(+51.14\%)}_{-0.019(-11.30\%)}$  \\ 
		$[500.00,540.00]$ 	 & 	  ${0.706}^{+0.078(+11.00\%)}_{-0.077(-10.96\%)}$ & 	 ${0.844}^{+0.113(+13.42\%)}_{-0.067(-7.98\%)}$  \\ 
		$[540.00,580.00]$ 	 & 	  ${1.164}^{+0.117(+10.06\%)}_{-0.123(-10.58\%)}$ & 	 ${1.238}^{+0.124(+10.06\%)}_{-0.111(-8.96\%)}$  \\ 
		$[580.00,620.00]$ 	 & 	  ${1.316}^{+0.121(+9.18\%)}_{-0.135(-10.25\%)}$ & 	 ${1.341}^{+0.112(+8.37\%)}_{-0.127(-9.47\%)}$  \\ 
		$[620.00,660.00]$ 	 & 	  ${1.280}^{+0.107(+8.38\%)}_{-0.127(-9.96\%)}$ & 	 ${1.271}^{+0.095(+7.51\%)}_{-0.117(-9.21\%)}$  \\ 
		$[660.00,700.00]$ 	 & 	  ${1.157}^{+0.088(+7.57\%)}_{-0.112(-9.68\%)}$ & 	 ${1.139}^{+0.073(+6.43\%)}_{-0.113(-9.89\%)}$  \\ 
		$[700.00,740.00]$ 	 & 	  ${1.008}^{+0.069(+6.84\%)}_{-0.095(-9.44\%)}$ & 	 ${0.984}^{+0.056(+5.69\%)}_{-0.091(-9.26\%)}$  \\ 
		$[740.00,780.00]$ 	 & 	  ${0.859}^{+0.052(+6.07\%)}_{-0.079(-9.19\%)}$ & 	 ${0.836}^{+0.039(+4.62\%)}_{-0.079(-9.43\%)}$  \\ 
		$[780.00,820.00]$ 	 & 	  ${0.722}^{+0.038(+5.27\%)}_{-0.064(-8.91\%)}$ & 	 ${0.705}^{+0.027(+3.76\%)}_{-0.065(-9.21\%)}$  \\ 
		$[820.00,860.00]$ 	 & 	  ${0.607}^{+0.028(+4.60\%)}_{-0.053(-8.72\%)}$ & 	 ${0.592}^{+0.018(+3.07\%)}_{-0.053(-8.93\%)}$  \\ 
		$[860.00,880.00]$ 	 & 	  ${0.530}^{+0.022(+4.15\%)}_{-0.046(-8.59\%)}$ & 	 ${0.515}^{+0.017(+3.31\%)}_{-0.044(-8.51\%)}$  \\ 
		$[880.00,980.00]$ 	 & 	  ${0.411}^{+0.013(+3.15\%)}_{-0.034(-8.27\%)}$ & 	 ${0.401}^{+0.007(+1.68\%)}_{-0.035(-8.73\%)}$  \\ 
		$[980.00,1100.00]$ 	 & 	  ${0.257}^{+0.004(+1.51\%)}_{-0.020(-7.87\%)}$ & 	 ${0.252}^{+0.000(+0.00\%)}_{-0.020(-8.11\%)}$  \\ 
		$[1100.00,1300.00]$ 	 & 	  ${0.137}^{+0.000(+0.00\%)}_{-0.010(-7.31\%)}$ & 	 ${0.135}^{+0.000(+0.00\%)}_{-0.010(-7.64\%)}$  \\ 
		$[1300.00,1600.00]$ 	 & 	  ${0.056}^{+0.000(+0.00\%)}_{-0.004(-6.61\%)}$ & 	 ${0.056}^{+0.000(+0.00\%)}_{-0.004(-6.95\%)}$  \\ 
		$[1600.00,1900.00]$ 	 & 	  ${0.020}^{+0.000(+0.00\%)}_{-0.002(-7.60\%)}$ & 	 ${0.021}^{+0.000(+0.00\%)}_{-0.002(-8.61\%)}$  \\  \bottomrule
	\end{tabular}
\end{table}

\clearpage



{\footnotesize
\bibliographystyle{JHEP}
\bibliography{bibliography}
}


\end{document}